\documentclass[12pt,fleqn,epsfig]{article}
\usepackage{graphicx}

\title{Individual-based model for coevolving competing populations }

\author{I. C. Charret$^1$, J.N.C. Louzada$^2$, A. T. Costa Jr.$^1$\\
$^1$Universidade Federal de Lavras,\\
Departamento de Ci\^encias Exatas, 37200-000 Lavras - MG\\
$^2$Universidade Federal de Lavras,\\ 
Departamento de Biologia, 37200-000 Lavras - MG}

\begin{document}

\maketitle

\begin{abstract}

Classical models for competition between two species usually predict 
exclusion or divergent evolution of resource exploitation. However, recent 
experimental data show that coexistence is possible for very similar species 
competing for the same resources \textit{without} niche partition. Motivated 
by this experimental challenge to classical competition theory,
we propose an individual-based stochastic competition model, which is
essentially a modification of a deterministic Lotka-Volterra type model.
The proposed model of competition dynamics incorporates the effects of a 
discrete genotype, which determines the individual's adaptation to the 
environment, as well as its interaction with the other species.\\


\end{abstract}

\section{Introduction}

Mathematical biology is a fast growing, well recognised area of research. 
Although not clearly 
defined yet, is consider one of the most exciting modern applications of 
mathematics. The increasing use of mathematics in biology is inevitable as 
biology becomes more quantitative. The complexity of the biological sciences 
makes interdisciplinary approach essential. In particular, mathematical 
modelling of the dynamics of populations interactions is a very challenging 
and important task. 

In building a mathematical description of interacting populations,  
the main goal is to capture the essential features of a complex 
trophic web with a mathematically tractable model. The result is,
usually, a non-linear dynamical system, which may be expressed
as a set of either differential or difference equations. Classical 
examples are the pioneering works of Lotka (1932) and Volterra (1931). 
Such models may be understood as describing the time 
evolution of population averages in a closed ecosystem. In this context, one 
can devise two broad categories of models: \textit{ad hoc} models, intended 
to give a phenomenological mathematical description of the dynamics, and 
\textit{ab initio} models, which try to stablish sets of dynamical rules 
based on or inspired by the biology of the populations. The majority of 
classical population models are deterministic. Recently, due to the growing 
availability of affordable computer power, \textit{individual-based models} 
have been becoming popular. Instead of describing only population sizes, 
they treat them as collections of individuals with distinctive 
characteristics. Some of those models have been successfull in tackling 
important biological problems, such as ageing and the evolution of sex (Penna, 1995) and (Oliveira, 1999). The possibility of combining elements of 
population 
dynamics and genetics into mathematical models is indeed exciting and 
promising. Individual-based models are, in general, stochastic models, 
known to have a rich variety of behaviors, and sometimes differing markedly 
from their deterministic counterparts. 

Classical models for competition between two species usually predict 
exclusion or divergent evolution of resource exploitation. These predictions
are massively corroborated by experience and observation. Such results are
commonly summarized in Gause's competitive exclusion principle: to coexist, 
species must differ in their resource use; otherwise one of them ends up 
extinct (Gause, 1935). However, recent experimental data show that coexistence 
is possible for very similar species competing for the same resources 
\textit{without} niche partition (Louzada, 1996).

Motivated by this experimental challenge to classical competition theory,
we propose an individual-based stochastic competition model, which is
essentially a modification of a deterministic Lotka-Volterra type model.
We introduce, as the individual characteristic, a rudimentary ``genotype'',
which determines the individual's adaptation to the environment, as
well as its interaction with the other species. In section II we describe the
model and its implementation. In section III we present the results of
computer simulations of the model proposed in section II, and discuss
their implications to coexistence in real ecosystems. We present our final 
remark and summarize the conclusions in section IV.

\section{A model for coevolving competing populations}

The continuous-time, deterministic Lotka-Volterra model for competition 
between two species is given by the equations

\begin{equation}
\frac{dN_i}{dt}= N_i\left(r_i - \alpha_{ij}N_j\right) - \alpha_{ii}N_i^2,
\,\,\, i,j=1,2,\,\,\, j\neq i\, .  \\
\label{classical_LV}
\end{equation}

where $N_i(t)$, $i=1,2$ are the populations sizes at time $t$, $r_i$, $i=1,2$ 
are the intrinsic growth rates; $\alpha_{ij}$, $i,j=1,2$, 
are the competition strengths. It is well known that equations 
\ref{classical_LV} have critical points 
$\vec{N}_{excl}^{(1)}=(r_1/\alpha_{11},0)$ and that
$\vec{N}_{excl}^{(2)}=(0,r_2/\alpha_{22})$ corresponds to exclusion of one 
species, while 
$\vec{N}_{coex}=(r_1\alpha_{22}-r_2\alpha_{12},r_2\alpha_{11}-r_1\alpha_{21})/
(\alpha_{11}\alpha_{22}-\alpha_{21}\alpha_{12})$ corresponds to coexistence. 
When $\alpha_{11}\alpha_{22}>\alpha_{12}\alpha_{21}$ the coexistence critical 
point is globally stable, whereas for 
$\alpha_{11}\alpha_{22}<\alpha_{12}\alpha_{21}$ the exclusion points are the 
globally stable ones, and coexistence becomes a saddle point. Put in biological
terms, whenever intra-specific competition is stronger than inter-specific
competition there can be coexistence; otherwise one species always exclude
the other.

We are interested in building a ``microscopic'' model for competition which
is able to mimic the most common biological observations, namely competitive 
exclusion and coexistence by character displacement. The model should also be 
able, given some specific conditions, to reproduce the less commonly observed
coexistence by \textit{converging} evolution. 

We start by converting equations \ref{classical_LV} into probabilistic rules
for individual reproduction/death as follows (May, 2001). The intra-specific
competition term, $\alpha_{ii}N_i^2$ is interpreted as a death process, giving 
a death probability $\alpha_{ii}N_i$. The probability that each individual 
of population $N_i$ reproduces succesfully is given by 
$r_i - \alpha_{ij}N_j$, $i\neq j$. 
This simple set of stochastic rules is more easily related to the discretized 
version of equations \ref{classical_LV},
\begin{equation}
N_i(t+1)= N_i(t) + N_i(t)\left(r_i - \alpha_{ij}N_j(t)\right) - 
\alpha_{ii}N_i(t)^2,\,\,\, i,j=1,2,\,\,\, j\neq i\, .  \\
\label{discrete_LV}
\end{equation}

Now time is a discrete variable, labeling generations, and the parameters
should be interpreted as being the ones in equations \ref{classical_LV} 
multiplied
by the time interval between generations. It is worth mentioning that the
discretized version \ref{discrete_LV} is identical to the continuous
version (equations \ref{classical_LV}) in the limit of very large populations.
Thus, it shares with the continuous model the critical points and their
stability properties, as can be easily seen by direct iteration 
of equation \ref{discrete_LV}.

Up to now the ``microscopic'' character of the model, meaning 
its description using rules for \textit{individual} birth/death 
processess, can be considered merely formal. It differs from
equation \ref{discrete_LV} only in a well-known demographic stochasticity 
\cite{May}, that is  relevant just in the limit of small populations. Our 
next step is to make the model really ``microscopic'', by endowing each 
individual with a property (which we call a genotype), that will modify 
competition and intrinsic growth. The phenotype will then be a measure 
of how close an individual's genotype is to the optimal genotype for a 
particular environment. The individual's genotype is also compared to 
the genotypes of individuals of the competing species. The closer they 
are, the stronger the competition is.

Genotypes $G_i[l]$ are 32-bit strings, associated with individual $l$
in population $i$. From each $G_i[l]$ two phenotypic traits are derived:
its normalized Hamming distance  to the environmentally determined optimal 
genotype $G_{opt}$, 
\[\tau_i[l]=\frac{\langle G_i[l],G_{opt}\rangle}{32}\, ,\] 
and its average normalized Hamming distance to the individuals of the 
opposing population, 
\[ \xi_i[l] = \frac{1}{N_j}\sum_{m=1}^{N_j}\frac{\langle G_i[l],G_j[m]\rangle}{32},\,\, j\neq i. \]
In order to simulate a harsh environment, we impose severe penalties to
individuals that depart from $G_{opt}$: their intrinsic reproduction probability
is assumed to decrease exponentially with $\tau_i[l]$ as
\[ r_i\exp(-\lambda\tau_i[l])\, , \]
where $\lambda$ is a parameter regulating the environment ``harshness''.
The competition will be modified according to

\[\alpha_{ij}(1-\xi_i[l]N_j),\,\, j\neq i\, .\]
which is a much milder dependence on the phenotype than that imposed
by the environment.

In order to allow the species to ``evolve'', at every birth event
the newborn inherits a copy of its mother's genotype with a possible
``mutation'', represented by a flip of a random bit. The possibility
of mutation introduces ``real'' (as opposed to the formal demographic) 
stochasticity into the model, and should alter its dynamical behaviour.
To investigate the dynamics we simulate the model in a computer.

\section{Simulation Results}

Our interest is to study the emergence of coexistence other than 
the ``trivial'' coexistence of the classical Lotka-Volterra 
model of equations \ref{classical_LV}. Thus we start with a situation
where intra-specific competition is weaker than inter-specific
competition. We also assume, for the time being,
symmetric competition ($\alpha_{12}=\alpha_{21}$), and identical initial
intrinsic reproduction rates ($r_1=r_2$). The genotypes of both 
populations are randomly distributed with uniform probability
over the space of 32-bit sequences. No special meaning is attributed 
to any particular sequence other than the optimal genotype $G_{opt}$.
For the situation just described, the only critical points of the cassical
model are $\vec{N}_{excl}$.

\begin{figure}
\centerline{\includegraphics[scale=0.5,clip]{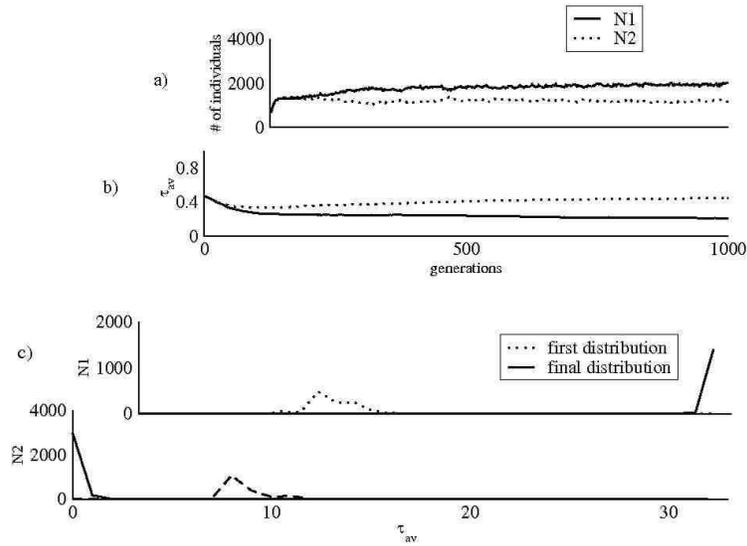}}
\caption{a) Predator and prey population sizes for initial conditions
, and parameters $r_1=r_2=0.8$, 
$\alpha_{11}=\alpha_{22}=0.5$,
$\alpha_{12}=\alpha_{21}=0.7$. 
The black and red curves are populations 1 and 2, respectively, with 
$N_1(0)=N_2(0)=500$. 
b) Average distance between populations' genotypes and $G_{opt}$, 
$\tau_{av}$, showing evolution towards 
niche partition. c) First and final distribution of genotypes showing 
divergent evolution.
(The simulation was run for a much longer time than shown to guarantee that 
the asymptotic regime had been reached).}
\label{fig1}
\end{figure}

Simulations of the proposed model show evolution towards coexistence,
as can be seen in figure \ref{fig1}. There, initial populations are
the same, $N_1(0)=N_2(0)=500$, but the asymptotic steady state is not 
symmetric. Due to random fluctuations, one of the species is forced to 
move away from $G_{opt}$, as can be seen in the inset, while the other 
approaches it as much as possible. This makes the average distance between 
the two populations genotypes very large, which minimizes the effects of 
competition. We may interpret this result as the system spontaneously 
attainning some kind of niche partition.

\begin{figure}
\centerline{\includegraphics[scale=0.5,clip]{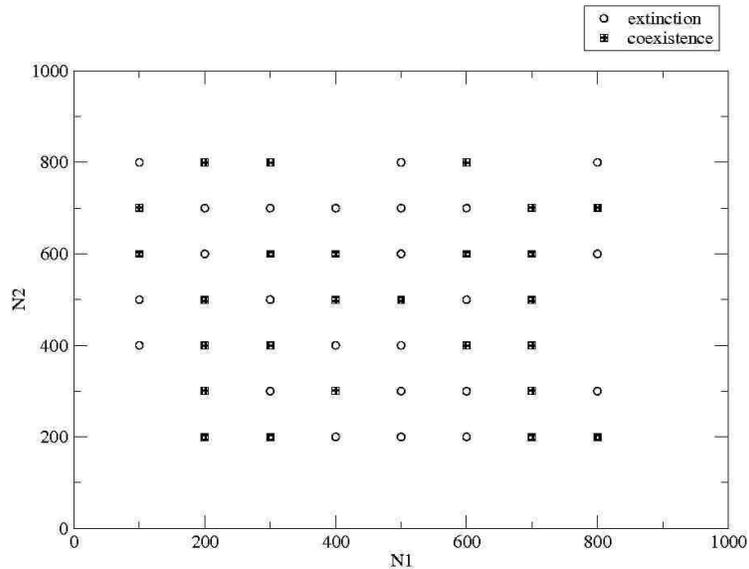}}
\caption{Sketch of the basin of attraction of the coexistence points
(squares) based on simulations with varying initial populations.
The squares are initial points from which coexistence is reached.}
\label{basin}
\end{figure}

For $N_1(0)=N_2(0)=800$ one of the species is excluded. This means that
the coexistence state (or states), contrary to the situation in the classical
Lotka-Volterra model, has limited basins of attraction, and
it is interesting to estimate the sizes of those basins. Starting with
$N_1(0)=1000,N_2(0)=600$, for example, the system reaches a steady state
with average populations different from the case of figure \ref{fig1}.
This is an indication of existence of mutliple coexistence critical 
points in this model. Nevertheless, all these coexistence states correspond 
to niche partition. In figure \ref{basin} we plotted
the initial points from which a coexistence state is reached for fixed
values of the parameters. As can be seen, the basin of attraction of the
stability region is rather limited, in contrast with the classical 
Lotka-Volterra coexistence point being globally stable.

\begin{figure}
\centerline{\includegraphics[scale=0.5,clip]{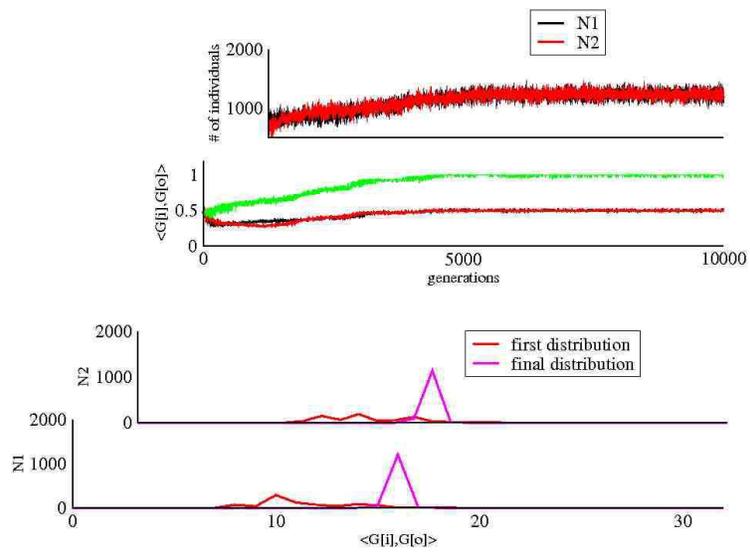}}
\caption{Coexistence with converging evolution. The asymptotic intra-specific
competition is larger than the inter-specific competition. a) Population sizes;
b) $\tau_{av}$. c) First and final distribution of genotypes showing convergent evolution.
(The simulation was run for a much longer time than shown to guarantee that 
the asymptotic regime had been reached).} 
\label{coex_conv}
\end{figure}

The simulations also showed cases of coexistence with converging evolution to 
a single phenotype, corresponding to a distance between individual's genotypes and $G_{opt}$, that allows for near-maximum variability inside populations. 
In these cases, exemplified by the results of figure \ref{coex_conv}, 
asymptotic inter-specific competition is weaker than
the asymptotic intra-specific competition. Once again, evolution leads to a
situation compatible with the classical Lotka-Volterra dynamics.

\begin{figure}
\centerline{\includegraphics[scale=0.5,clip]{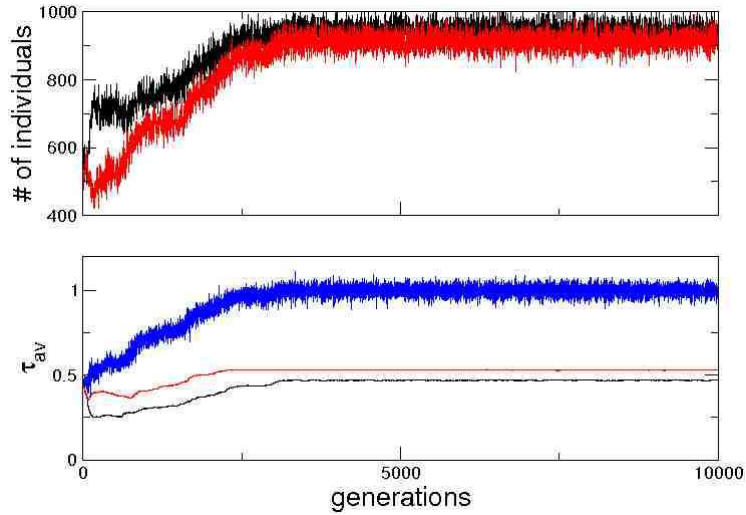}}
\caption{Coexistence with converging evolution. a) Population sizes
are very similar; b) notice that the average distances 
$\tau_{av}$, are both large. This is compatible with large initial 
variability inside each
population, what allows for a large average distance between populations'
genotypes (blue curve).}
\label{coex_aleat}
\end{figure}

Another interesting case of coexistence has very similar steady-state
population sizes and also similar values of $\langle\tau_i\rangle$, as
seen in figure \ref{coex_aleat}. However, the average genotypic distance 
between populations may be high. This is indeed what happens in this 
particular case, as can be seen in Fig. \ref{coex_aleat}b. Once again, 
the system found its way through evolving towards minimal competition, 
and coexistence is thus possible.

\section{Concluding remarks}

\subsection{Qualitative comparison to experiments}

The results provided by the proposed model show coexistence attained
through minimization of competition. This is compatible with the
expectations based on classical population models. However, the observations
that motivated this work are not exactly contemplated by our results.

The two species of Dichotomius studied by Louzada \textit{et al.} show 
identical patterns of utilization of and preference for resources. Besides,
they show identical abundance in all experiments performed to date. Hence, 
resource sharing can not explain the coexistence of the two species. 
The fact that the two species exploit identically the temporal, spatial 
and feeding axes of their niche support the need for a more elaborate 
explanation of coexistence in Scharabaeidae communities. It may be thought that
the mechanism through which these communities minimize competition is
related to a sudden change of competitive behaviour: there seems to be
a similarity threshold beyond which individuals of different species
start to identify each other as pertaining to the same species, and their
competition turns from the inter-specific to the intra-specific type.
This mechanism is completely absent from the model proposed in this work.
We are investigating possible modifications of the proposed model to
take this kind of mechanism into account.

\subsection{Summary of results}

We presented an individual-based model for the dynamics of two coevolving
competing species. It is an extension of the Lotka-Volterra equations for
competition to include coevolution. The dynamical behaviour of the proposed 
model shows important differences from the classical Lotka-Volterra 
competition, 
such as multiple coexistence critical points with limited attraction basins.
The coevolution dynamics mimics important biological behaviours such as
niche partition and converging evolution. Modifications of the present model
to reproduce other kinds of experimentally observed behaviour are underway.

{\bf Acknowledgments}

We gratefully acknowledge  J. S. S. Bueno Filho for many 
enlightening  discussions, important suggestions and a critical reading of 
the manuscript. A.T.C. acknowledges financial support from CNPq. I.C.C. and 
J.N.C.L. acknowledges financial support from FAPEMIG.

\end{document}